\documentclass[10pt,twocolumn,letterpaper]{article}

\usepackage{multicol,lipsum}
\usepackage{cuted}
\usepackage{flushend}
\usepackage{graphicx}
\usepackage{amsmath}
\usepackage{amssymb}
\usepackage{booktabs}

\usepackage{amsmath,amsfonts,bm}

\def\eqref#1{equation~\ref{#1}}

\def\1{\bm{1}}

\DeclareMathAlphabet{\mathsfit}{\encodingdefault}{\sfdefault}{m}{sl}
\SetMathAlphabet{\mathsfit}{bold}{\encodingdefault}{\sfdefault}{bx}{n}

\usepackage{hyperref}
\usepackage{url}
\usepackage{comment}
\usepackage{graphicx}
\usepackage{graphics}
\usepackage{booktabs}
\usepackage{multirow}
\usepackage{tabu}
\usepackage{bbding}
\usepackage{pifont}
\usepackage{amsmath}

\newcommand\on{\mathit{on}}
\newcommand\off{\mathit{off}}

\newcommand{\yes}[0]{\footnotesize \Checkmark }
\newcommand{\no}[0]{\scriptsize \XSolidBrush}
\newcommand{\commentout}[1]{}
\usepackage{xspace}
\newcommand{\tool}{{\sc AntidoteRT}\xspace}

\newcommand{\lab}[1]{\textquotesingle{#1}\textquotesingle}
\usepackage{graphicx}

\usepackage{algorithm2e}

\usepackage[capitalize]{cleveref}
\crefname{section}{Sec.}{Secs.}
\Crefname{section}{Section}{Sections}
\Crefname{table}{Table}{Tables}
\crefname{table}{Tab.}{Tabs.}

\def\cvprPaperID{*****} 
\def\confName{XXX}
\def\confYear{2022}

\begin{document}

\title{AntidoteRT: Run-time Detection and Correction of Poison Attacks on Neural Networks}

\author{Muhammad Usman\\
University of Texas at Austin\\
{\tt\small muhammadusman@utexas.edu}
\and
Youcheng Sun\\
Queen's University Belfast\\
{\tt\small youcheng.sun@qub.ac.uk}
\and
Divya Gopinath\\
KBR Inc., NASA Ames\\
{\tt\small divgml@gmail.com}
\and
Corina S. P\u{a}s\u{a}reanu\\
Carnegie Mellon University, KBR, NASA Ames\\
{\tt\small pcorina@cmu.edu}
}
\maketitle

\begin{abstract}
We study backdoor poisoning attacks against image classification networks, whereby an attacker inserts a trigger into a subset of the training data, in such a way that at test time, this trigger causes the classifier to predict some target class. 
We propose lightweight automated detection and correction techniques against poisoning attacks, which are based on neuron patterns mined from the network using a small set of clean and poisoned test samples with known labels. The patterns built based on the mis-classified samples are used for run-time detection of new poisoned inputs. For correction, we propose an  input correction technique that uses a differential analysis to identify the trigger in the detected poisoned images, which is then reset to a neutral color. Our detection and correction are performed at run-time and input level, which is in contrast to most existing work that is focused on offline model-level defenses.
We demonstrate that our technique outperforms existing defenses such as NeuralCleanse and  STRIP on popular benchmarks such as MNIST, CIFAR-10, and GTSRB against the popular BadNets attack and the more complex DFST attack.
\end{abstract}

\section{Introduction}

\begin{figure*}
\centering
\includegraphics[scale=0.5]{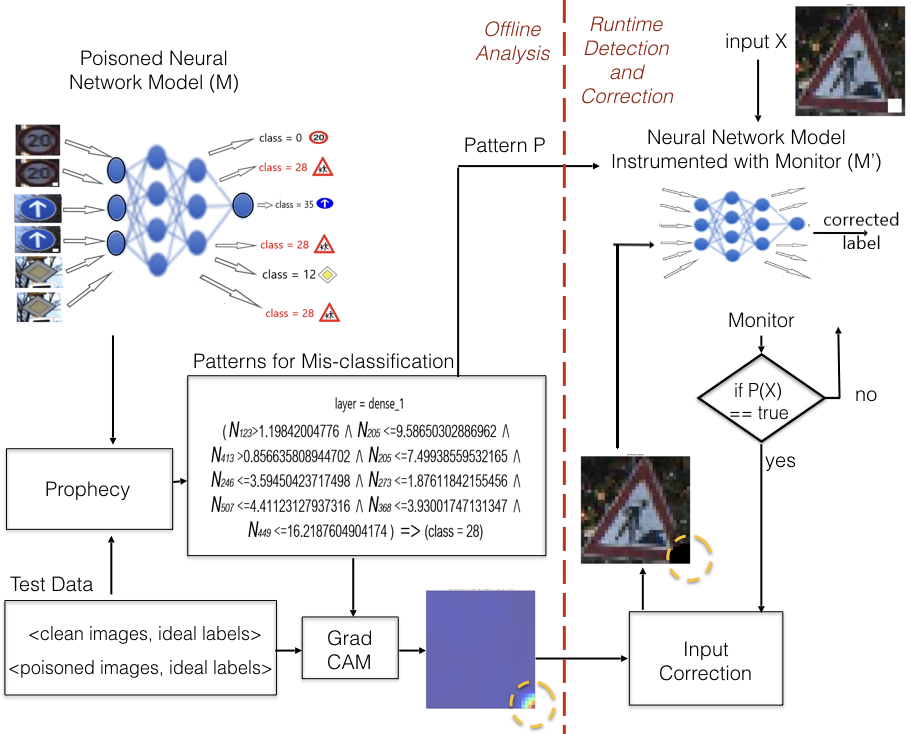}
\caption{Overview of \tool Analysis Framework.}
\label{fig:overview}
\end{figure*}

Neural networks have been increasingly used in a variety of safety-related applications \cite{huang2020survey}, raging from manufacturing, medical diagnosis to perception in autonomous driving, and thus affecting many aspects of modern life. There is thus a critical need for techniques and tools to ensure that neural networks work as expected and are free of bugs and vulnerabilities. 

In this work, we propose lightweight, run-time detection and mitigation techniques against poisoning attacks on neural network image classifiers. These are well known attacks \cite{gu2019badnets,trojaning,cheng2021deep} that  are concerned with a malicious agent inserting a {\em trigger} into a subset of the training data, in such a way that at test time, this trigger causes the classifier to (wrongly) predict some target class.
There are a few techniques proposed in the literature that aim to detect the attack at run-time e.g., \cite{steinhardt2017certified,wang2019neural} but 
existing defense work \cite{10.1007/978-3-030-00470-5_13,yao2019latent,li2020rethinking,liu2020reflection} typically involves retraining/finetuning the network which is still not feasible for run-time use. 


Previous work \cite{ASE19} proposed the use of decision tree learning to extract {\em likely properties} of neural networks, based on observing the neuron activations (on/off) at intermediate layers. These properties are {\em likely} in the sense that they are satisfied by a large number of inputs and they can be formally proved using existing verification tools. In this work we propose a generalization of the previous work \cite{ASE19} to also consider the neuron values when computing the patterns and explore the application of the generalized patterns for the detection and correction of poisoning attacks.

\noindent{\bf Threat Model.}
Given a trained model, our assumption is that we have a test data set that can be used for assessing the model. While the train set is not required, our approach does assume that the test set contains a small percentage of poisoned inputs with known ground truth labels. 
This corresponds to a typical software testing scenario where the user observes anomalies during testing of a software component and aims to debug and remedy the problem.
Note that our technique does not need to know if the model is poisoned or the poisoned target label. However, poisoned models have high accuracy on clean (un-poisoned) data, so most of the mis-classified inputs are likely poisoned. 

\noindent{\bf Approach.}
We propose to extract neural network patterns in terms of mathematical constraints over the neuron values at intermediate layers in such a way that they {\em discriminate} between correct and incorrect classifications of the network. Thus if a new input satisfies the constraints for a mis-classification pattern it will likely be mis-classified. 
We show that these patterns 
can serve as good {\em run-time detectors} of poisoned inputs. We further propose a mitigation strategy that first identifies the {\em trigger} in the poisoned inputs by using a differential analysis based on off-the-shelf attribution techniques; we use GradCam \cite{chattopadhay2018grad}
to identify the input pixels that most likely influence the neurons in the patterns. The inputs are then {\em corrected} (at run-time) by setting the identified pixels to a neutral color. 

 

We first evaluate our proposed detection and correction technique in the context of a backdoor attack that inserts a small patch as the trigger~\cite{gu2019badnets} for the well known MNIST, CIFAR-10 and GTSRB data sets; furthermore, we evaluate a more challenging deep feature space trojan (DFST) attack that was proposed recently \cite{cheng2021deep} that evades state-of-the-art defenses. 
We show experimentally that our technique significantly outperforms existing approaches NeuralCleanse \cite{wang2019neural} and  STRIP \cite{gao2019strip}, thus promising to be a viable detection and repair technique against poison attacks at run-time. 
%
%
We summarize our contributions as follows: {\em i)} Novel run-time detection technique for poisoned inputs based on neuron patterns, {\em ii)} Novel input correction technique based on a differential analysis, in contrast to altering the logic of the network (such as \cite{wang2019neural}), {\em iii)} Novel idea to use patterns to guess the ideal label for poisoned inputs at run-time, as demonstrated on the more challenging DFST attack, {\em iv)} Experimental evidence showing that \tool has better detection and repair rates than state-of-the-art techniques.
\section{Background}
\label{sec:background}

\noindent{\bf Prophecy.}
\label{subsec:prophecy}
Prophecy~\cite{ASE19} is a tool which extracts decision patterns for a neural network model $F$ based on a set of data; the goal is to build assume-guarantee style rules with respect to an output property of the model. 
A decision pattern $\sigma$ is a pre-condition for network $F$ with respect to a post-condition $P$ if: 
$\forall X: \sigma(X) \Rightarrow P(F(X))$.
In the case of classifier models the  post-condition is that the output class is equal to a certain label. The rule is in terms of neuron activation patterns, that is, 
neurons ($N$) at intermediate layer/s being on or off. 
$$\sigma(X) := \bigwedge_{N \in \on(\sigma)} N(X) > 0  \wedge  \bigwedge_{N \in \off(\sigma)} N(X) \leq 0$$ 
Prophecy extracts these patterns by applying decision tree learning over the activation signatures recorded for the given labeled data when executed on the model. Each pattern can be proven using an off-the-shelf verification tool such as Marabou~\cite{katz2019marabou}. However, a pattern can be used as a detector without providing a formal proof. Each such pattern is associated with a \textit{support}, which indicates the number of data instances that satisfy the rule. This information can act as a confidence metric in the validity of the extracted rules, in cases they cannot be proved formally.

\noindent{\bf GradCAM++.}
\label{subsec:gradcam}
GradCAM++~\cite{chattopadhay2018grad} is a recent approach for explaining the decisions of convolutional neural network models used for image classification. It aims to generate class activation maps that highlight pixels of the image input that the model uses to make the classification decision. It builds on the idea proposed in~\cite{selvaraju2017grad} of using the gradients of any target concept flowing into the final convolutional layer to produce a coarse localization map highlighting the important regions in the image for the model to make a prediction. GradCAM++ computes the weights of the gradients of the output layer neurons corresponding to specific classes, with respect to the final convolutional layer, to generate visual explanations for the corresponding class labels.
\section{Approach}
\label{sec:approach}

Figure~\ref{fig:overview} presents an overview of our \tool approach, which has two major components: (1) {\em offline analysis} and (2) {\em run-time detection and correction}. 
\tool takes as inputs a trained image classifier and a small set of test (instead of training) data along with their ground-truth labels. \tool executes the model on the data and learns if the inputs are correctly classified (passing inputs) or not (failing inputs). 
Given a model that has been poisoned during training, it has normal accuracy on clean data (i.e., inputs without the backdoor trigger) and it mis-classifies the poisoned inputs to a certain target label.  
For the example in Figure~\ref{fig:overview}, a GTSRB (German Traffic Sign Recognition Benchmark \cite{Houben-IJCNN-2013}) model is attacked, and images embedded with a white patch at the bottom right would be mis-classified to the target label 28.
\tool has a realistic setup. It does not  need the knowledge of whether the model has been poisoned at training, or what the poison target label is.
Moreover, \tool does not require that its input clean test data is the corresponding original data  for the poisoned ones used by it. 

\subsection{Offline Analysis}
\label{sec:offline}
The offline analysis in \tool comprises of two parts: extracting mis-classified patterns (Section \ref{subsec:pattern_extraction}) and localizing the input trigger (Section \ref{subsec:iden_pixels})

\subsubsection{Extraction of mis-classification patterns}
\label{subsec:pattern_extraction}

A backdoor attack on neural network models involves poisoning a set of images with some backdoor trigger during the training phase, which introduces an incorrect logic in the model whereby it learns to recognize the backdoor trigger as a dominant feature that it associates with the poison target label. Our goal is to determine this incorrect logic 
in a poisoned model. 
To this end, we modify the Prophecy tool (Section~\ref{subsec:prophecy}) to extract  {\em mis-classification patterns}. 

A mis-classification pattern is a rule at a certain layer of the network that distinguishes or discriminates the behavior of the model for correctly classified and mis-classified inputs. 
Patterns for incorrect behavior potentially represent the incorrect logic introduced by the poisoning, and are thus good candidates for detecting new poisoned inputs at run-time. The example pattern in Figure~\ref{fig:overview} 
is a rule in terms of the output values of neurons at the dense layer just before the output layer. The rule indicates that inputs satisfying this pattern have a high chance of being mis-classified to  output class 28, that is the target label from the attack. 

 In the past, Prophecy has been applied to extract patterns in terms of neuron activations, where the threshold for a neuron activation (typically ReLU) was zero. However the positive neuron outputs themselves can vary in a wide range of values, which could in turn impact the model outputs. Therefore, we modified Prophecy to feed the actual neuron values to the decision-tree learner, such that a suitable threshold may be selected for each neuron as part of learning the tree for the different labels. 

Specifically, the inputs are executed on the model and the output values of the neurons at the dense layer closest to the output are recorded. In our work, we always choose the {\em last dense layer}, before the output layer. The justification is as follows. Typically layers closer to the input layer (such as the convolutional layers) focus on input processing for feature extraction, while the dense layers close to the output hold the logic in terms of the extracted features which impact the network's output. 
A data set is created with the neuron values recorded for each input. 
Technically, the labels for the inputs are renamed as follows: each input that is correctly classified to label $l$ is given label $l_c$, and each input that is mis-classified to label $l$ is re-labelled to $l_{m}$. Decision-tree learning is then invoked to extract rules at the dense layer for the re-named labels.


Prophecy is thus used to extract the following rules.

$$\forall X\;\;\sigma^{l}_{c}(X) \Rightarrow (F(X) = l \wedge l = l_{ideal})$$
$$\forall X\;\;\sigma^{l}_{m}(X) \Rightarrow (F(X) = l \wedge l \neq l_{ideal})$$





Here $\sigma^{l}_{c}(X)$ represents a pattern for correct classification to label $l$, $\sigma^{l}_{m}(X)$ represents a pattern for mis-classification to $l$.
Both patterns have this form: 
$$ \bigwedge_{N_{i} \in {\mathcal N}_{L}} N_{i}(X) \; op \; V_{i}.$$
${\mathcal N}_{L}$ is the set of neurons at layer $L$, 
$V_{i}$ is the threshold value for the output of neuron $N_{i}$ (as computed by decision tree learning), $op$ is the operator in $\{ >, <=\}$, and $F(X)$ is the output of the model.
%
Note that Prophecy extracts {\em pure} rules, i.e., all inputs satisfying a pattern lead to same label.

Since for poisoned models, most of the mis-classifications are due to the poisoned data, we can determine that the label corresponding to the mis-classification pattern with the highest support  should be the \textit{poison target label}.
We select all patterns for mis-classification to the poison target label 
to form the set $\mathcal{P}$, which is used for detection at run-time. 



\subsubsection{Identification of trigger pixels in the image}
\label{subsec:iden_pixels}
Poisoned images have the backdoor trigger embedded within the input image. The second step of the offline analysis is to identify the portion of the image corresponding to the trigger. 
We propose a differential analysis over images that are correctly and incorrectly classified; this analysis uses off-the-shelf attribution techniques to identify the pixels of the input impacting the poison target label (class 28 in the example). 
In our work we use GradCAM++ (Section~\ref{subsec:gradcam}) for attribution but other techniques can be used as well.
%
The output is a heatmap that identifies the portions of the input for the backdoor trigger. As shown in Figure~\ref{fig:overview} the heatmap highlights pixels in the bottom right of the image, corresponding to the location of the white patch in the poisoned inputs in our example. 



In order to identify the image pixels corresponding to the trigger we define a differential analysis over heatmaps computed with gradient attribution approaches such as GradCAM++. Typically gradient attribution approaches identify pixels of an input image that impact the model output. However, applying such a technique on a per input basis may lead to a heatmap that is over fitted to the specific input and may also not be precise depending on the noise in the input image. Therefore, a per-image attribution is not useful in identifying the vulnerable portions of the image that is generalizable to unseen inputs. Given that the mis-classification patterns potentially capture the incorrect logic of the model in terms of input features, we utilize them to group inputs that satisfy the same pattern and obtain a summary of the important pixels across the images. 

There can be more than one mis-classification patterns in $\mathcal{P}$. 
For every such mis-classification pattern $p\in\mathcal{P}$, 
a summarized heatmap $HM_{p}$ is created. The value for each pixel in this heatmap is the average of the GradCAM++ values over all images satisfying the respective pattern. 

$$\forall p \in {\mathcal P}\;\;HM_{p} =  \sum_{X \in {\mathcal X}_{p}} GradCAM(X) / \# {\mathcal X}_{p},$$
$\;\;\;\;\;\;\;\;\;\;\;\; \forall X\;\;X \in {\mathcal X}_{p} \iff p(X) = True$

Our differential analysis is drawing inspiration from traditional fault localization approaches 
that use both passing tests and failing tests to isolate the fault inducing entity. Similarly  we use the attribution for images corresponding to correct classification as well to obtain a more precise localization of the pixels that comprise the features that cause incorrect behavior. Thus we create a summarized heatmap for all correctly classified images as follows.

$$HM_c =  \sum_{X \in {\mathcal X}_{c}} GradCAM(X) / \# {\mathcal X}_{c},$$
$\;\;\;\;\;\;\;\;\;\;\;\; \forall X\;\;X \in {\mathcal X}_{c} \iff F(X) = l_{ideal}$ 

The heatmaps for the poisoned inputs 
are then normalized as follows.

$${HM'_{p}[i]} = HM_{p}[i] / \sum_{j}{HM_{p}[j]}$$


We normalize $HM_c$ to create $HM'_c$ in a similar manner.
We then create a difference heatmap ($\Delta_p$) per pattern to better isolate the pixels impacting the incorrect behavior. Each pixel has a value that is the difference of its value in the heatmap corresponding to the poisoned inputs satisfying a pattern and its value in the heatmap corresponding to correct inputs. 

$$\forall p \in {\mathcal P}\;\;\Delta_{p} = HM'_p - HM'_c
$$

The value of each pixel in $\Delta_{p}$ is representative of its impact specifically on the incorrect behavior of the model. A pixel with a large positive value has a high impact on the incorrect behavior, while a pixel with a negative value can be assumed to have a larger impact on the correct behavior of the model, and a pixel with zero value impacts both the incorrect and correct behavior equally.  

We short-list the top \textit{threshold} \% of the total number of pixels based on their values in $\Delta_{p}$ to form the set of important pixels for the respective pattern ($pix_p$).  $Imp\_Pixels$ is the set of $pix_p$ for all patterns in ${\mathcal P}$, $Imp\_Pixels = \{ pix_p \}$. This is fed to the run-time module to enable run-time correction. 

\subsection{Run-time detection and correction}

\SetKwComment{Comment}{/* }{ */}
\SetKwFor{For}{for (}{) }{$\textbf{end}$}

\begin{algorithm}

\caption{Run-time detection and input repair.
$F$ is the original classifier function, $\mathcal{P}$ is the list of mis-classification patterns  sorted in descending order of support, $Imp\_Pixels$ is also sorted to match the order of patterns in $\mathcal{P}$.}

\label{alg:pseudocode}

\KwData{$X$, $F$, ${\mathcal P}$, $Imp\_Pixels$}
\KwResult{$label$}

/*\textit{\textbf{detection}}*/

$found \gets False$;

$indx \gets 0$;

\While{$indx \leq \#{\mathcal P}$}{
  \If{$indx =  \#{\mathcal P}$}{
  
      /*\textit{\textbf{no match, not poisoned, orig label}}*/
      $label \gets F(X)$;
   }
   $p \gets {\mathcal P}[indx]$;
   
   \If{$p(X)$ $=$ True}{
   
      $found \gets True$;
      
      \textbf{break};
   }
   
   $indx \gets indx + 1$;
}

/*\textit{\textbf{input-based correction}}*/

\If{$found$ = True}{

  $pix \gets Imp\_Pixels[indx]$;

  $X' \gets X$;

  /*\textit{\textbf{mask pixels}}*/
  
  \For{$j \in pix$}{
      $x'[j] \gets 0$;
  }

  $label \gets F(x')$;
}
\end{algorithm}

The outputs of the offline analysis, the mis-classification patterns ${\mathcal P}$  and the important pixels  for every mis-classification pattern, are fed as inputs to the run-time module. Algorithm~\ref{alg:pseudocode} gives the pseudo-code for run-time detection and correction.

${\mathcal P}$  is used to create a run-time monitor to detect potentially poisoned inputs at run-time (forming $M'$ in Figure~\ref{fig:overview}). For any new input ($X$) at run-time, the monitor checks if the neuron values at the respective intermediate layer satisfy the conditions of any of the mis-classification pattern 
in ${\mathcal P}$. If the check fails, the input is considered to be not poisoned, and the original label is produced by the model. However, if the check passes, the input is considered to be poisoned. The processing of the input is stopped and the input image is passed onto a correction module. 

Note that the patterns are mutually-exclusive only on the input data used to extract them in the offline analysis. A new input at run-time could satisfy more than one patterns in ${\mathcal P}$. In such a case, we would conclude that the input is potentially poisoned, however, we would need to choose one of the satisfying patterns to pass on to the next step of correction. We choose the pattern with the highest support since patterns with high support are more likely to be valid (refer Section~\ref{sec:background}). Thus (with slight abuse of notation), in Algorithm~\ref{alg:pseudocode} we assume ${\mathcal P}$ is a list of patterns sorted in descending order of support. Similarly, we assume ${Imp\_Pixels}$ represents a list of important pixels for every pattern in the same order as in ${\mathcal P}$.
Please refer the algorithm steps under \textit{detection} in Algorithm~\ref{alg:pseudocode}.

The important pixels for the mis-classification pattern chosen in the previous step are obtained from $Imp\_Pixels$. These are used to correct the input image. The portion of the image corresponding to the important pixels are masked to remove the trigger.  The \textit{masking} that our tool supports currently is setting the pixel values to a neutral value (such as zero). This works well on the benchmarks that we have analyzed (refer Section \ref{sec:eval}). 
We plan on investigating more sophisticated techniques for input correction as future work (refer Section~\ref{sec:future}). The model is invoked again on the modified input $x'$ and the corresponding (corrected) label is produced.  

The example in Figure~\ref{fig:overview} shows that the road sign image with the backdoor trigger gets incorrectly classified by the poisoned model as label 28. However, the poisoned input gets detected by the monitor at the dense layer as its signature at that layer satisfies the pattern shown in the figure. The correction step masks the trigger to produce the modified image shown. This input when fed back into the model, gets classified correctly to label 25.

\section{Evaluation}
\label{sec:eval}

In this section, we evaluate the performance of \tool, by comparing it with three baselines based on two attack techniques and three datasets. 

\subsection{Attack techniques and baselines} 

\begin{figure}[!ht]
\centering
  \includegraphics[]{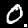}
  \includegraphics[]{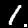}
  \hspace{0.3cm}
  \includegraphics[]{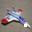}
  \includegraphics[]{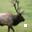}
  \hspace{0.3cm}
  \includegraphics[]{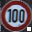}
  \includegraphics[]{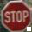}
  \caption{Example poisoned data for MNIST (left), CIFAR-10 (middle) and GTSRB (right). The backdoor is embedded as the white square at the bottom right side of each image. When the backdoor appears, the poisoned MNIST model will classify the input as {\lab 7}, the poisoned
  CIFAR-10 model will classify it as {\lab {horse}} and the poisoned GTSRB model will classify the input as {\lab{watch for children}}.}
  \label{fig:badnets_samples}
\end{figure}

\begin{figure}[!ht]
\centering
  \includegraphics[]{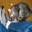}
  \includegraphics[]{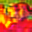}
  \hspace{.3cm}
  \includegraphics[]{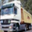}
  \includegraphics[]{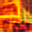}
  \caption{Poisoned examples from DFST: Each pair is a clean input and its DFST poisoning counterpart}
  \label{fig:dfst_samples}
\end{figure}

Two representative attack techniques BadNets \cite{gu2019badnets} and DFST \cite{cheng2021deep} are used for evaluating \tool.
\begin{itemize}
    \item \textbf{Fixed trigger for all inputs.}
BadNets is the most common type of backdoor trigger to neural network models, wherein attack techniques have fixed pixel-space patches, watermarks or color patterns as the trojan trigger. Figure \ref{fig:badnets_samples} shows how the BadNets attack embeds the trigger on the three data sets in the evaluation, and Table \ref{tab:badnets-models} reports the poisoned models' performance on clean data and poisoned data respectively.    
\item \textbf{Different triggers for different inputs.}
DFST is the latest backdoor attack technique wherein the features of the backdoor trigger are different at the pixel level for different inputs. They are injected into the benign inputs through a specially trained generative model called trigger generator. We use this technique to poison the CIFAR-10 model such that the trigger is the sunset style (Table \ref{tab:dfst-models}). Figure \ref{fig:dfst_samples} shows the normal inputs and poisoned inputs by DFST side by side.
\end{itemize}

\begin{table}[t]
\scalebox{0.85}{
\begin{tabular}{cccc}\toprule
Dataset & \begin{tabular}{c}Clean\\ Accuracy\end{tabular} & \begin{tabular}{c}Attack \\Success Rate\end{tabular} & \begin{tabular}{c}Model\\ Architecture\end{tabular}\\\hline
MNIST & 98.95\% & 97.94\%& 2 conv + 2 dense\\
CIFAR10 & 82.24\% & 94.36\% &4 conv + 2 dense\\
GTSRB & 96.29\% & 97.24\% &6 conv + 2 dense\\\bottomrule
\end{tabular}
}
\caption{Poisoned models via BadNets: Clean accuracy is the poisoned model's performance on clean validation data and attack success rate measures how effective the attack is when the test validation data is poisoned.}
\label{tab:badnets-models}
\end{table}

\begin{table}[t]
\centering
\scalebox{0.85}{
\begin{tabular}{ccc}\toprule
Dataset & \begin{tabular}{c}Clean\\ Accuracy\end{tabular} & \begin{tabular}{c}Attack \\Success Rate\end{tabular} \\\hline
CIFAR10 & 81.70\% & 99.66\% \\\bottomrule
\end{tabular}
}
\caption{Poisoned models via DFST}
\label{tab:dfst-models}
\end{table}

The three baselines for run-time detection/correction are from STRIP \cite{gao2019strip} and NeuralCleanse \cite{wang2019neural}. Though there is a significant body of work on backdoor attack/defense of neural networks, when it comes to \emph{run-time} detection and correction, these baselines are regarded as the state of the art. 

STRIP is a technique focusing on detecting poisoned inputs for a poisoned model. Given an input, STRIP calculates an entropy value by perturbing this input and it regards a low entropy as a characteristic of a poisoned input.

\begin{figure}
\centering
  \includegraphics[]{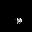}
  \includegraphics[]{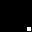}
  \caption{The  synthesized trigger by NeuralCleanse (left) and the groundtruth trigger for GTSRB}
  \label{fig:nc_triggers}
\end{figure}

NeuralCleanse synthesizes a potential trigger for each output label and calculates an anomaly measure from them to decide if some label was the target of backdoor attack. Its input detection and repair is based on the neuron activation values from the synthesized trigger, the higher the value is the more important the neuron is for identifying and removing the backdoor. We further designed another baseline that enhances the performance of NeuralCleanse, by feeding the groundtruth backdoor trigger to its detection/correction algorithm (\textit{NeuralCleanse (Groundtruth)}). Figure \ref{fig:nc_triggers} shows that the synthesized trigger by NeuralCleanse could be different from the groundtruth trigger. As we are going to see in the experiments, this difference impacts the detection/correction rates. 
The DFST attack does not have a fixed trigger for all inputs and so the groundtruth trigger enhancement for NeuralCleanse does not apply.

\subsection{Experiment setup}

\begin{table*}
	\centering
	\caption{Summary of results with BadNets attacks. \tool: \textit{threshold} set to customized \% based on GEN set.}
\label{tab:summary1}
\scalebox{0.8}{	\begin{tabular}{|c|l|c|c|c|c|c|}
			\hline
			\textbf{Tool} & \textbf{Metric (\%)} & \multicolumn{1}{c|}{\textbf{MNIST}} &\textbf{CIFAR-10} & \textbf{GTSRB} \\ \hline

\multirow{4}{*}{\tool (25\%)} & Poisoned Detection Rate  &\textbf{86.48}    &83.14  &\textbf{95.43}  \\ 
                      & Poisoned Repair Rate & 85.76   &\textbf{62.25}  &\textbf{93.54}  \\ 
                      & Clean Detection Rate &99.10    &98.28  &99.50  \\ 
                      & Clean Repair Rate & \textbf{98.57}   &\textbf{81.93}  &\textbf{96.32}  \\ \cline{2-5}
 \hline

\multirow{4}{*}{\tool (1\%)} & Poisoned Detection Rate  &54.24    &75.79  & 91.79\\ 
                      & Poisoned Repair Rate & 30.99   &47.81  & 70.64\\
                      & Clean Detection Rate & 99.38   &98.49  &99.72 \\
                      & Clean Repair Rate & 98.54   &82.16  & 96.27\\ \cline{2-5}
 \hline

\multirow{4}{*}{STRIP} & Poisoned Detection Rate  &54.31&\textbf{89.10} &0.00  \\ 
                      & Poisoned Repair Rate &N/A    &N/A  &N/A  \\ 
                      & Clean Detection Rate &\textbf{99.78}    &98.27  &\textbf{100.00}  \\ 
                      & Clean Repair Rate & N/A   &N/A  &N/A  \\ \hline
               
\multirow{4}{*}{NeuralCleanse} & Poisoned Detection Rate  & 79.69   & - & 0.00 \\ 
                      & Poisoned Repair Rate &\textbf{90.90}    &-  &3.77  \\ 
                      & Clean Detection Rate &99.65    &-  &\textbf{100.00}  \\ 
                      & Clean Repair Rate &94.62&-    &93.74   \\ \hline

\multirow{4}{*}{NeuralCleanse (Groundtruth)} & Poisoned Detection Rate  &83.35    &53.46  &86.49  \\ 
                      & Poisoned Repair Rate &90.29    &18.96  & 16.87 \\ 
                      & Clean Detection Rate &86.83    & \textbf{100.00} &\textbf{100.00}  \\ 
                      & Clean Repair Rate &91.92    & 77.35 & 95.67 \\ \hline

\hline
\end{tabular}
}
\end{table*}

We evaluate \tool on three vision benchmarks: MNIST \cite{lecun1998gradient}, CIFAR-10 \cite{krizhevsky2009learning} and GTSRB \cite{Houben-IJCNN-2013}. For each benchmark, we have a clean test set and poisoned test set respectively. We use these to create two types of datasets for each benchmark; a \emph{Generation (GEN)} set which is the input data in the offline analysis and a held-out \emph{Validation (VAL)} set for evaluating the performance of the tool at run-time. \emph{VAL} contains 50\% of the clean test inputs and 50\% of poisoned inputs randomly selected from the respective test sets. The rest of the clean test inputs and a percentage $\alpha$, varying across $\{1\%, 5\%, 10\%, 25\%\}$, of the remaining poisoned inputs are used to create \emph{GEN} sets. For instance, for MNIST, we generated 10k poisoned inputs corresponding to 10k clean test inputs, and used them to create a VAL set with 5k clean and 5k poisoned inputs, and GEN sets with 5k clean inputs and poisoned inputs ranging from 50 (1\%) to 1250 (25\%) respectively (refer supplementary material for details). 
The random input selection process ensures that the GEN sets do not contain the respective correct versions for every poisoned input.

To analyze the detection/correction results, we use four complementary metrics: 1) poisoned detection rate, 2) poisoned repair rate, 3) clean detection rate and 4) clean repair rate. The clean (poisoned) detection rate is the percentage of clean (poisoned) inputs that are correctly detected at run-time, and clean (poisoned) repair rate measures the model's accuracy on the clean (poisoned) inputs after correction.  All experiments are repeated 10 times and average results are reported. We experiment with four different values for the \emph{threshold} to select the important pixels for input-based correction (Section~\ref{sec:approach}). We set it to 2\%, 5\%, 10\% and to a customized value based on the performance on the respective \emph{GEN} set. The code and data used are publicly available\footnote{\url{https://github.com/AntidoteRT/AntidoteRT}}.

\subsection{Results on benchmarks with BadNets attacks}
Table \ref{tab:summary1} summarizes the results from our tool, \tool, with two $\alpha$ parameters and the baselines STRIP, NeuralCleanse (NC) and NeuralCleanse with Groundtruth Trigger (NCGT) on the BadNets benchmarks (Table \ref{tab:badnets-models}). 
STRIP only has detection capabilities and cannot perform repair. Thus, we only report its clean and poisoned detection rates. 

Here is a summary of the general observations. Firstly, we can see that \tool is more effective than other baselines with respect to the detection and correction of poisoned data in most cases (refer \textit{Poisoned Detection} and \textit{Poisoned Repair} rates under the three benchmarks), with few false positives (refer \textit{Clean Detection} rates). Secondly, given that \tool only attempts to perform correction on inputs detected as being poisoned, our approach impacts the performance on clean data lesser than NeuralCleanse (refer \textit{Clean Repair rates}). Note that being a run-time technique, \tool does not perform any permanent change to the model. Overall, \tool gives good rates in a stable manner for all three benchmarks (unlike the other approaches), and specifically performs well on the more complex benchmark (GTSRB).  





\noindent \textbf{GTSRB}:
\tool successfully detects 95.43\% of the poisoned test inputs, which is significantly better than STRIP (0\%), NC (0\%) and NCGT (86.49\%). Meanwhile, input-based correction fixes 93.54\% of the total poisoned test inputs, which is much higher than NeuralCleanse even with the groundtruth trigger enhancement (16.87\%). \tool detects slightly less clean test inputs (99.50\% that is still very high) than others: STRIP (100.00\%), NC (100.00\%) and NCGT (100.00\%). 
\tool's repair has the least negative impact on the model's performance on clean data; this is demonstrated by the higher clean repair rate by \tool (96.32\%) than NeuralCleanse (93.74\%) and its enhancement (95.67\%). In fact, after the repair, only \tool increases the model's performance on clean data that is originally 96.29\%.


\noindent \textbf{CIFAR-10}:
\tool has 83.14\% poisoned detection rate which is slightly lower than STRIP (89.10\%) but significantly better than NCGT (53.46\%), whereas it has a better clean detection rate (98.28\%) than STRIP (98.27\%) but lower than NCGT (100.00\%). However, the repair performance of \tool is much better than NCGT on both poisoned (62.25\% vs 18.96\%) and clean (81.93\% vs 77.35\%) test data. Note that NC identified a wrong target label for the CIFAR-10 model and failed in detecting and repairing any poisoned input. 
{\tool outperforms baseline tools in the repair metrics and strikes a balance between STRIP and NCGT on the detection tasks.}

\noindent \textbf{MNIST}:
\tool achieves the best poisoned detection rate (86.48\%) and clean repair rate (98.57\%); its clean detection rate is very close to the best performed on STRIP (99.10\% vs 99.78\%). The original version of NC corrects the most poisoned inputs after its model repair (90.90\%), that is even better than NCGT which assumes the presence of the groundtruth trigger (86.83\%).


\noindent \textbf{Effect of Reducing the value of $\alpha$}: Table \ref{tab:summary1} also summarizes results of experiments with $\alpha$ = 25\% and the extreme case of $\alpha$ = 1\%, representing the percentage of poisoned test data used in the offline analysis of \tool. We can observe that the clean detection/repair rates change very slightly (~$1\%$) for the different $\alpha$ values. 
Poisoned detection and repair rates decreased more significantly. For detection: MNIST (86.48\% to 54.24\%), CIFAR-10 (83.14\% to 75.79\%) and GTSRB (95.43\% to 91.79\%); for repair: MNIST (85.76\% to 30.99\%), CIFAR-10 (62.25\% to 47.81\%) and GTSRB (93.54\% to 70.64\%). However, note that the rates for the 1\% setting are still better than both the other techniques for GTSRB, better than NCGT for CIFAR and STRIP for MNIST. 

Refer supplementary material for more detailed results. For any benchmark, \tool consumed a maximum of around 23 minutes for the offline analysis and had a runtime overhead of around 0.07 seconds per input. 

\subsection{Case Study on the DFST attack}
\label{sec-dfst}

\begin{table}
	\centering
	\caption{DFST sunrise attack on CIFAR-10 model.}
\label{tab:DFST}
\scalebox{0.8}{
		\begin{tabular}{|c|l|c|c|c|}
			\hline
			\textbf{Tool} & \textbf{Metric (\%)} & \multicolumn{1}{c|}{\textbf{CIFAR-10}} \\ \hline

\multirow{4}{*}{\tool} & Poisoned Detection Rate  & \textbf{88.26}   \\ 
                      & Poisoned Repair Rate (input-based) & 4.6 \\ 
                      & Poisoned Repair Rate (patterns-based) & \textbf{20.62}  \\ 
                      & Clean Detection Rate & 97.19  \\ 
                      & Clean Repair Rate & \textbf{80.64}   \\ \hline
                      
\multirow{4}{*}{STRIP} & Poisoned Detection Rate  &  0 \\ 
                      & Poisoned Repair Rate &  NA \\ 
                      & Clean Detection Rate & 98.45 \\ 
                      & Clean Repair Rate &  NA \\ \hline

\multirow{4}{*}{NeuralCleanse} & Poisoned Detection Rate  & 6.11  \\ 
                      & Poisoned Repair Rate &  10.02 \\ 
                      & Clean Detection Rate & \textbf{99.8} \\ 
                      & Clean Repair Rate & 78.07 \\ \hline

\hline
\end{tabular}
}

\end{table}

The DFST attack performs a more complex transformation of the inputs. Therefore we describe the application of \tool on it as a separate case study. 
The run-time 
detection using mis-classification patterns performed well. As can be seen in Table~\ref{tab:DFST}, \tool  detects more than 88\% of poisoned inputs correctly, which is much higher than the other techniques. However, the input-based correction did not work well for this attack with a poisoned repair rate of only 4\%. 
We surmised that our input-based correction is too simple for this attack, considering that the image is transformed in a non-trivial manner.


For correction, we experimented with a novel lightweight approach to \textit{guess} the ideal label of a potentially poisoned input at run-time. The idea is to build new patterns for correct classification, where the poisoned inputs are given ideal label instead of actual label. Thus, a pattern for label $l$ groups together un-poisoned inputs that are correctly classified to $l$ together with poisoned inputs that are mis-classified but have ideal label $l$. Intuitively, the new patterns are trained to both recognize the poisoned inputs and to output the correct label. To build these patterns, we run decision-tree learning once more using clean inputs for different labels and re-labeling poisoned inputs to their ideal labels.




A list of such patterns, $\mathcal P_{c}$, sorted in descending order of support is fed to the run-time module in addition to $\mathcal P$. At run-time, when an input is detected as being poisoned (based on $\mathcal P$), the execution is stopped and the input is checked against the patterns in $\mathcal P_{c}$. The label corresponding to the first pattern in $\mathcal P_{c}$ that is satisfied by the input is \textit{guessed} as the potential correct label for the input. If there is no pattern in $\mathcal P_{c}$ that is satisfied by the input, then a random label (excluding the poison target label) is assigned to the input. As shown in Table~\ref{tab:DFST} we were able to obtain a poisoned repair rate of {\bf 20.26\%}, which is much better than NeuralCleanse (10\%) and STRIP (0\%). 



\section{Related Work}
\label{sec:related_work}
Existing poison detection techniques provide methods either for detecting individual backdoor inputs at run-time (similar to ours, NeuralCleanse and STRIP) or for identifying the backdoor model (most existing work). 
As we already described NeuralCleanse and STRIP in evaluation, we discuss briefly here model detection techniques.

\noindent{\textbf{Model Detection.}} Backdoor detection techniques such as \cite{steinhardt2017certified,turner2018clean,liu2017neural,tran2018spectral} rely on statistical analysis of the poisoned training dataset for deciding if a model is poisoned or trojaned. In \cite{chen2019detecting}, it is shown that activations of the last hidden neural network layer for clean and legitimate data and the activations for backdoor inputs form two distinct clusters. 
DeepInspect \cite{chen2019deepinspect} learns the probability distribution of potential triggers from the queried model using a conditional GAN model, which can be used for inspecting whether the pre-trained DNN has been trojaned.
Kolouri et al. \cite{kolouri2020universal} pre-define a set of input patterns that can reveal backdoor attacks,
classifying the network as `clean' or `corrupted'. 
The TND (TrojanNet Detector) in \cite{wang2020practical} explores connections between Trojan attack and prediction-evasion adversarial attacks.
In \cite{xu2021detecting}, a meta-classifier is trained that predicts whether a model is backdoored.




\noindent{\textbf{Correction.}}
Different from the input correction method developed in this paper, existing defense techniques on neural network backdoor are focusing on re-training, fine-tuning  or pruning \cite{10.1007/978-3-030-00470-5_13,yao2019latent,li2020rethinking,liu2020reflection,liu2021removing,li2020neural}. 
These works end up with the fundamental and difficult neural network parameter selection problem, for effectively erasing the impact of backdoor triggers from the model without degrading (much) the model's overall performance. In contrast, with our technique, the effect on already correctly classified inputs is minimal.
The work in \cite{udeshi2019model} is the only other input-level repair that we are aware. Unlike our technique, it is black box and therefore much more expensive. 
It repeatedly searches the area of an image for the position of the backdoor trigger, which is accomplished by placing a trigger blocker of the dominant colour in the image. 

\section{Conclusion, Limitations and Future Work}
\label{sec:future}

We presented lightweight run-time  detection and correction techniques against poisoning attacks, that are based on neuron patterns mined from the network. We demonstrated that \tool outperforms  existing  defenses on popular vision benchmarks against two types of attacks. It produces stable results across benchmarks, with small run-time overhead. 

One limitation is that our technique requires a set of poisoned inputs, albeit very small. In contrast, STRIP and NeuralCleanse do not require any prior poisoned data. We showed experimentally that the small amount of poisoned data used by \tool enables consistently better detection rates for the poisoned inputs. Furthermore, \tool outperforms NeuralCleanse when it was even `helped' with the groundtruth backdoor trigger, demonstrating the value of using activation patterns for detection and correction. 

The input repair by \tool works against simple backdoor attacks, but not against the more complex DFST attack, where using patterns for guessing correct labels worked better.  Future work involves employing machine learning to infer a more complex repair that can function as a reliable antidote at the input level. We also plan to perform more experiments with the idea of using patterns for guessing the right label.

{\small
\bibliographystyle{ieee_fullname}
\bibliography{main}
}


\end{document}